\documentclass[aps,prl,preprint,groupedaddress]{revtex4-2}
\usepackage{color}
\usepackage{graphicx}
\usepackage{amsmath}

\begin{document}

%\preprint{}

\title{Conversion of stable crystals to metastable crystals in a solution by periodic change of temperature}


\author{Hiroyasu Katsuno}
\email[]{katsuno@lowtem.hokudai.ac.jp}
%\thanks{}
\affiliation{Institute of Low Temperature Science, Hokkaido University, Kita-19, Nishi-8, Kita-ku, Sapporo, Hokkaido, 060-0819, Japan}

\author{Makio Uwaha}
\affiliation{Science Division, Center for General Education, Aichi Institute of Technology, 1247 Yachigusa, Yakusa-cho, Toyota, Aichi 470-0392, Japan}


\date{\today}

\begin{abstract}
Using a Becker-D\"oring type model
 including cluster incorporation,
 we study the possibility of conversion of stable crystals to metastable crystals in a solution by a periodic change of temperature.
At low temperature, both stable and metastable crystals grow by coalescence with abundant clusters.
At high temperature,
 a large amount of small clusters produced by the dissolution of crystals inhibits the dissolution of crystals,
 and the imbalance in the amount of crystals increases.
By repeating this process, 
 the periodic temperature change can convert stable crystals into metastable crystals.

\end{abstract}

%\keywords{}

\maketitle

%\section{}
%\subsection{}
%\subsubsection{}

\section{Introduction}

Among many possible crystal structures,
 the most stable structure is realized and metastable structures disappear in equilibrium.
Exceptional cases are chiral crystals which have two stable structures that are
 thermodynamically
equivalent.
When the most stable structure is chiral, 
 it is inevitable that both left- and right-handed crystals exist in equilibrium.
In 2005, 
 Viedma demonstrated
 the conversion of a racemic mixture of chiral sodium chlorate crystals into homochiral crystals by grinding crystals in a solution\cite{viedma2005chiral}.
The chirality conversion by grinding is also possible in conglomerate forming chiral molecules that racemize in a solution\cite{noorduin2008emergence,sogutoglu2015viedma}.
In contrast with Ostwald ripening observed in the process of relaxation to equilibrium,
 it is called Viedma ripening (VR).
Several years later, 
 some researchers showed a similar phenomenon with periodic temperature change (temperature cycling: TC) of a solution.
The boiling of a solution with powder crystals is a simple way to realize TC \cite{el2009spontaneous,viedma2011homochirality}.
More quantitative experiments were performed with controlled temperature change of a solution\cite{suwannasang2013using,suwannasang2016novel}.
One of the important features of VR and TC is the exponential amplification of an initial small crystal enantiomeric excess during the conversion.

From the theoretical analysis\cite{saito2004complete} of chemical reactions that bring complete homochirality, 
 it is likely that important factors to realize a homochiral state in VR/TC are also nonlinear autocatalysis and recycling of the product.
Nonlinear autocatalysis increases the asymmetry in the ratio of two enantiomers
 and the product is recycled to produce the dominant enantiomer.
In VR/TC experiment,
 the enhancement of dissolution of crystals by grinding crystals/increasing temperature corresponds to the recycling process.
Then, the question is what processes in crystallization correspond to the nonlinear autocatalytic process.
Several mechanisms for VR 
have been proposed such as
 the chiral cluster incorporation in crystallization\cite{uwaha2004model}, 
 the catalytic surface reaction\cite{saito2008chiral},
 the mutual inhibition\cite{saito2010crystal} based on the Frank model\cite{frank1953spontaneous},
 and the secondary nucleation by the shear stress of the flow\cite{cartwright2007ostwald}.
The chiral cluster incorporation is adopted for the explanation of VR in various approaches:
 the simple rate equation\cite{uwaha2004model,uwaha2008simple,mcbride2008did,steendam2015linear,spix2016persistent,katsuno2017effect},
 the time change of the crystal size distribution
 such as a Backer-D\"oring (BD) model\cite{uwaha2009mechanism,wattis2011mathematical,iggland2011population,blanco2013viedma,blanco2015modeling}
 and a population balance (PB) model\cite{iggland2014effect},
 and Monte Carlo simulation\cite{katsuno2009monte,ricci2013computational}.
The cluster incorporation mechanism developed in BD type models also explains the chirality conversion by TC
 for achiral and chiral molecules\cite{blanco2013viedma,katsuno2016mechanism}.
For the crystallization of chiral molecules,
 simple molecular incorporation without clusters
 described by a PB model reproduces the chirality conversion in TC\cite{bodak2018population,bodak2019effect}.
There is no common view on the mechanism of TC at the moment\cite{uwaha2022mechanism}.

While the essential process in TC for chirality conversion is still not very clear,
 the chirality conversion by TC is completely different from the relaxation to equilibrium.
The results of TC can be interpreted as the system chooses one state from two thermodynamically equivalent states.
The final state is determined by the initial state and dynamics of the system.
Here, a simple question arises: is it possible to choose an energetically unfavorable state from two thermodynamically inequivalent states in TC?
A recent experiment with the combination of VR and TC shows an extraordinary phenomenon
 that stable racemic crystals are converted to metastable chiral crystals\cite{viedma2021new}.
In a numerical study,
 the conversion to the metastable phase crystals by VR is confirmed\cite{katsuno2021possibility}.
These results show that the conversion of the phase of crystals (hereafter, we call phase conversion) is possible by the simple methods,
 and that even metastable phase crystals win if the growth rate of metastable crystals becomes larger than that of stable phase crystals.

In this paper,
 we study the phase conversion by TC in a solution which contains the stable and the metastable phase crystals.
We discuss the mechanism of the phase conversion in detail.

\section{Model}

\begin{figure}[t]
\centering
\includegraphics[width=\linewidth]{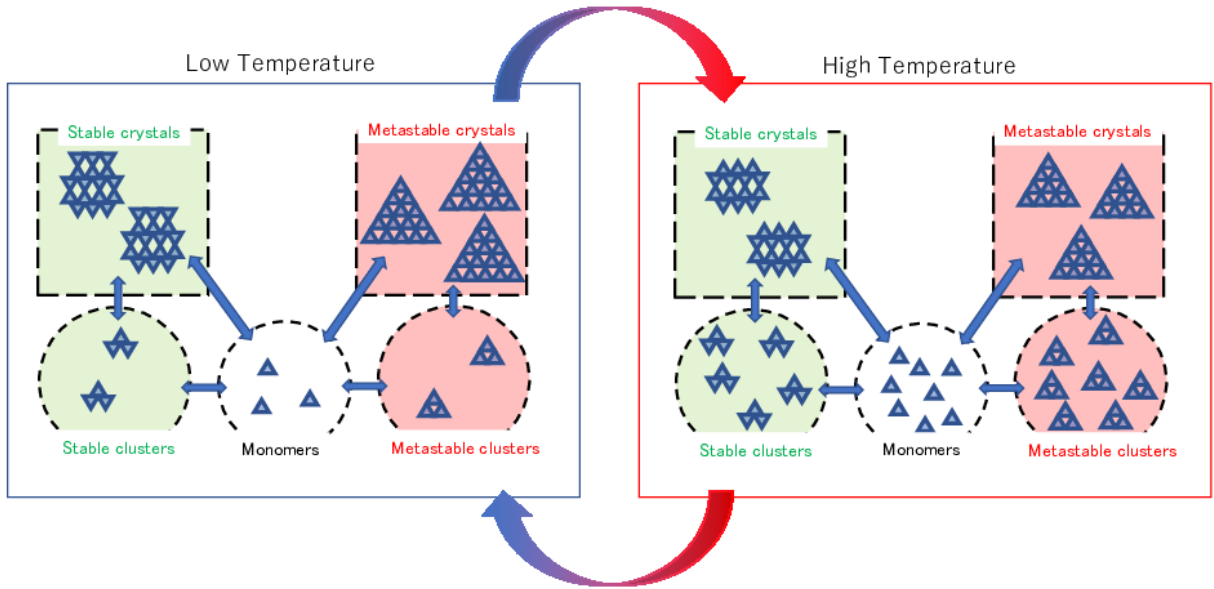}
\caption{
Schematics of our model.
``Crystals'' means clusters whose size is larger than $i_{\rm s}$ and do not coagulate with other ``crystals''.
}
\label{fig:figs/figs_p2}
\end{figure}

For the investigation of the phase conversion by TC,
 we use a generalized BD model with incorporation of clusters in crystallization.
The original BD model describes the change of the cluster size distribution 
 through monomers\cite{becker1935r,markov2016crystal,uwaha2010classical}.
Our generalization of the BD model are 
 i) introduction of the stable and the metastable crystal phases
 and ii) cluster incorporation into crystals of the same phase as schematically shown in Fig.~\ref{fig:figs/figs_p2}.
In our model,
 the stable and metastable phases are characterized by the two parameters: the equilibrium number
 of monomers (solubility)\footnote{
In the original BD model,
 the time change of the number density is described.
 Here, we call the number density the number for simplicity.
}
  and the interfacial energy.
Although two sets of parameters are necessary for the stable and the metastable phases,
 we assume the interfacial energies of both phases are the same.
This setting ensures that the metastable phase crystals are always energetically unfavorable for any given size.
The solubilities of the stable phase $n_{1, \rm s}^{\rm eq}$ and the metastable phase $n_{1,\rm m}^{\rm eq}$ are related to the difference in the chemical potential:
$\Delta \mu_{\rm m} - \Delta \mu_{\rm s} = k_{\rm B}T\ln\left(n_{\rm 1, s}^{\rm eq}/n_{\rm 1, m}^{\rm eq}\right)<0$,
 where $\Delta \mu_{\rm \gamma}$ is the difference in chemical potential between the $\gamma$-phase crystal and the solution,
 $k_{\rm B}$ is the Boltzmann constant,
 and $T$ is temperature.
Hereafter, we call $n_{1, \rm s}^{\rm eq}/n_{1,\rm m}^{\rm eq}(<1)$ the solubility ratio.

For the generalization ii),
 the acceptable incorporation process depends on the cluster size $i$.
Small clusters, whose size is smaller than $i_{\rm s}$, can be incorporated into clusters/crystals (we call small clusters simply clusters).
Large clusters, whose size is larger than $i_{\rm s}$, can not be incorporated into other crystals (we call large clusters simply crystals) .

In the numerical calculation,
 we investigate the time change of the number $n_{i,\gamma}$, where $\gamma=\textrm{m}$ and $\gamma=\textrm{s}$ represent the metastable phase and the stable phase.
The change of $n_{i,\gamma}$ by incorporation of a monomer or other clusters is given by
 $\sigma_{i,j}n_{i,\gamma}n_{j,\gamma}$,
 where the incorporation rate $\sigma_{i,j}$ is proportional to the collision cross-section of the size $i$ and $j$, {\it i.e.}, $\sigma_{i,j}=ai^{2/3}j^{2/3}$ with a coefficient $a$.
The change by dissolution is $\lambda_{i,j}^{\gamma} n_{i, \gamma}$,
 where the dissolution rate $\lambda_{i,j}^{\gamma}$ of $j$-mer from $i$-mer is determined from the detailed balance condition:
\begin{align}
\lambda_{i,j}^{\gamma}n_{i,\gamma}^{\rm eq}=\sigma_{i-j,j}n_{i-j,\gamma}^{\rm eq}n_{j,\gamma}^{\rm eq},
\end{align}
where $n_{i,\gamma}^{\rm eq}$ is the equilibrium Boltzmann distribution
 and we have assumed that $\sigma_{i,j}$ is independent of $\gamma$ for simplicity.
The Boltzmann distribution is given by the solubility $n_{1,\gamma}^{\rm eq}$ and the free energy of an $i$-mer
\begin{align}
n_{i,\gamma}^{\rm eq} = n_{1,\gamma}^{\rm eq}\exp[-\bar{\alpha}(i^{2/3}-1)],
\end{align}
where $\bar{\alpha}$ corresponds to the interfacial energy divided by $k_{\rm B}T$.
Change of the number of monomers by the incorporation/dissolution process is written as 
\begin{align}
\frac{\partial n_{1}}{\partial t}
 = -2\sigma_{1,1}(n_{1})^{2}
   +\sum_{\gamma= {\rm s}, {\rm m}}\left[
       - \sum_{j=2}^{i_{\rm max}-1} \sigma_{1,j}n_{1}n_{j,\gamma} + 2\lambda_{2,1}^{\gamma}n_{2,\gamma} + \sum_{j=2}^{i_{\max}-1}\lambda_{j+1,j}^{\gamma}n_{j+1,\gamma}
   \right],
\end{align}
The number of monomers is denoted by $n_{1}( = n_{1,\rm m} = n_{1,\rm s})$ because monomers are common.
For simplicity,
 it is assumed that a dimer formed by the coalescence of monomers belongs to the stable or the metastable phase with the equal probability.
Change of the number of clusters with $2 \le i \le i_{\rm s}$ is written as 
\begin{align}
\frac{\partial n_{i, \gamma}}{\partial t}
 = \sum_{j=1}^{[(i+1)/2]} \left(\sigma_{i-j,j}^{\prime}n_{i-j,\gamma}n_{j, \gamma} - \lambda_{i,j}^{\gamma}n_{i,\gamma}\right)
   -\sum_{j=1}^{i_{\rm max}-i} \left[ \left(\sigma_{i,j}+\delta_{i,j}\sigma_{i,i}\right)n_{i,\gamma}n_{j,\gamma} -  \lambda_{i+j,j}^{\gamma}n_{i+j,\gamma}\right],
\end{align}
where $\sigma_{i,j}^{\prime}$ is defined as 
\begin{align}
 \sigma_{i,j}^{\prime} =
\left\{
\begin{array}{cc}
 \sigma_{1,1}/2 \hspace{2em}& \textrm{if}\ \ i=j=1 \\
 \sigma_{i,j}   \hspace{2em}& \textrm{otherwise},
\end{array}
\right.
\end{align}
 $\delta_{i,j}$ is the Kronecker delta, 
and $i_{\rm max}$ is the maximum size of crystals preset in our numerical calculation.
Change of the number of crystals with $i > i_{\rm s}$ is 
\begin{align}
\frac{\partial n_{i,\gamma}}{\partial t}
 = \sum_{j=1}^{\textrm{min}\{i_{\rm s},[(i+1)/2]\}} \left(\sigma_{i-j,j}n_{i-j,\gamma}n_{j,\gamma} - \lambda_{i,j}^{\gamma}n_{i,\gamma}\right)
   -\sum_{j=1}^{i_{\rm s},i+j \leq i_{\rm max}}\left(\sigma_{i,j}n_{i,\gamma}n_{j,\gamma}-\lambda_{i+j,j}^{\gamma}n_{i+j,\gamma}\right).
\end{align}

We periodically change temperature $T$ of the system.
The temperature profile is simple: $T$ is constant and low in the first half of the period $P$ and high in the second half.
The solubility $n_{1,\gamma}^{\rm eq}$ at low temperature is smaller than that at high temperature.
The effective interfacial energy $\bar{\alpha}$
 at low temperature is larger than $\bar{\alpha}$ at high temperature.
A weak temperature dependence of the constant $a$ in the incorporation rate is ignored.

The parameters in our numerical calculation are as follows:
The constant in the incorporation rate is $a=1$.
In the low temperature,
 the solubility of the metastable crystals is $n_{\rm 1, m}^{\rm eq}=10^{-2}$
 and the effective interfacial energy is $\bar{\alpha}=5$.
In the high temperature,
 $n_{\rm 1,m}^{\rm eq}=1.5\times 10^{-2}$ and $\bar{\alpha}=0.5$.
The period of the temperature cycle is $P=10$.
The total mass is conserved and normalized: $n_{1}+\sum_{i>1,\gamma}in_{i, \gamma}=1$.
The maximum crystal size is $i_{\rm max}=1000$,
 and the maximum cluster size is $i_{\rm s}=10$.
The initial condition is that all monomers are in crystals and they have the same size $i=500$.

\section{Numerical results \& discussion}
The state of the system is indicated by the relative crystal mass difference between the stable and the metastable crystals
\begin{align}
\phi = \frac{M^{\rm m} - M^{\rm s}}{M^{\rm m} + M^{\rm s}},
\end{align}
where the mass of $\gamma$-phase crystals is defined by 
\begin{align}
M^{\rm \gamma} = \sum_{i>i_{\rm s}}^{i_{\rm max}} in_{i,\gamma}.
\label{eqn:mass}
\end{align}
We call $\phi$ the excess parameter in the present paper.

\begin{figure}[t]
\centering
\includegraphics[width=0.45\linewidth]{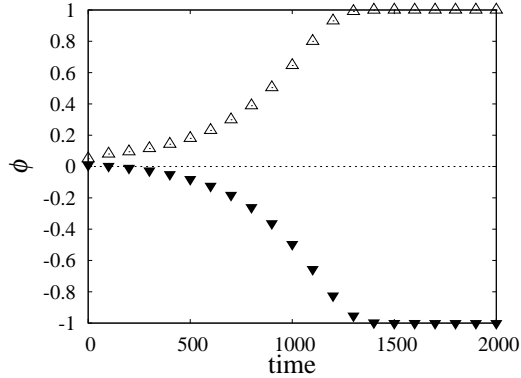}
\caption{
Time change of the excess parameter $\phi$.
Open triangle: $\phi(0)=0.05$, filled inverted triangle: $\phi(0)=0.01$.
The dotted line shows $\phi=0$.
}
\label{fig:fortran/BDmodel/12temp/test12/param4/omega-diffneq/0.99/init/0.05/ee}
\end{figure}

We demonstrate the conversion of the stable crystals to the metastable crystals by TC
 when the difference in solubility is small.
The time change of the excess parameter $\phi$ with the solubility ratio $n_{\rm 1, \rm s}^{\rm eq} /n_{\rm 1,m}^{\rm eq}=0.99$ is shown in Fig.~\ref{fig:fortran/BDmodel/12temp/test12/param4/omega-diffneq/0.99/init/0.05/ee}.
When the initial condition is $\phi(0)=0.05$ (open triangle in Fig.~\ref{fig:fortran/BDmodel/12temp/test12/param4/omega-diffneq/0.99/init/0.05/ee}),
 all stable crystals are converted to metastable crystals.
When the initial condition is $\phi(0)=0.01$ (filled inverted triangle in Fig.~\ref{fig:fortran/BDmodel/12temp/test12/param4/omega-diffneq/0.99/init/0.05/ee}),
 all metastable crystals are converted to the stable crystals.

In the previous work\cite{katsuno2021possibility},
 it was shown that the phase conversion of the stable crystals to the metastable crystals is possible by VR.
Using similar parameter values of the solubilities and the interfacial energies (see Fig.~4 in \cite{katsuno2021possibility}),
 the conversion by TC is faster than that by VR.
In an experiment of the chirality conversion, 
 the conversion time of TC is about $20$ times shorter than that of VR\cite{belletti2022combining}.
Our result is consistent with the experimental result
 although direct correspondence between the theoretical and the experimental parameter values cannot be made.

In our result shown in Fig.~\ref{fig:fortran/BDmodel/12temp/test12/param4/omega-diffneq/0.99/init/0.05/ee},
 the initial excess parameter $\phi(0)$ is amplified exponentially in time.
The behavior of the excess parameter is similar to that of the crystal enantiomeric excess of chiral crystals during VR and TC.
However, the metastable phase crystals can win only when $\phi(0)$ is above a critical value.

\begin{figure}[t]
\centering
\includegraphics[width=0.32\textwidth,clip]{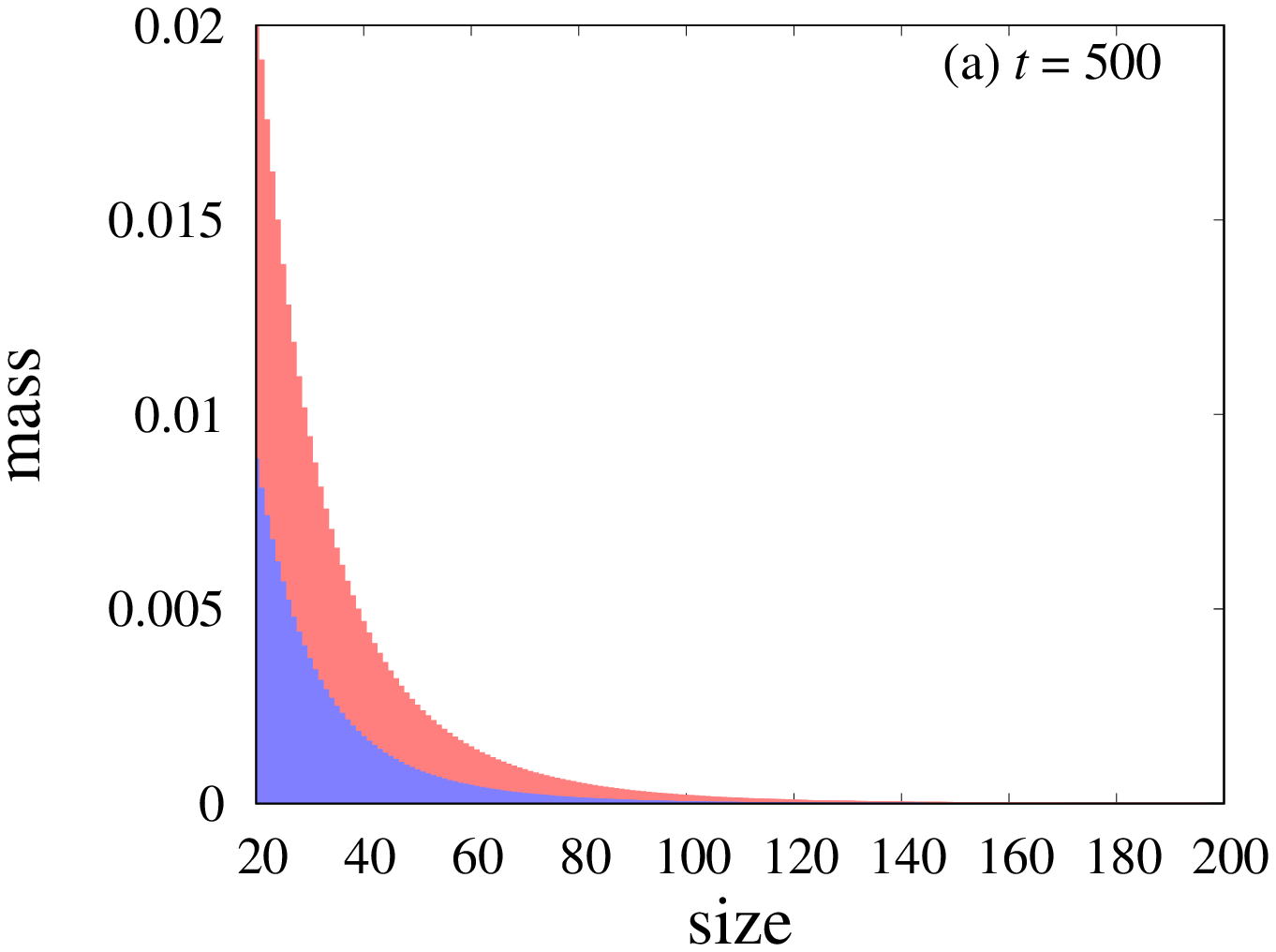}
\includegraphics[width=0.32\textwidth,clip]{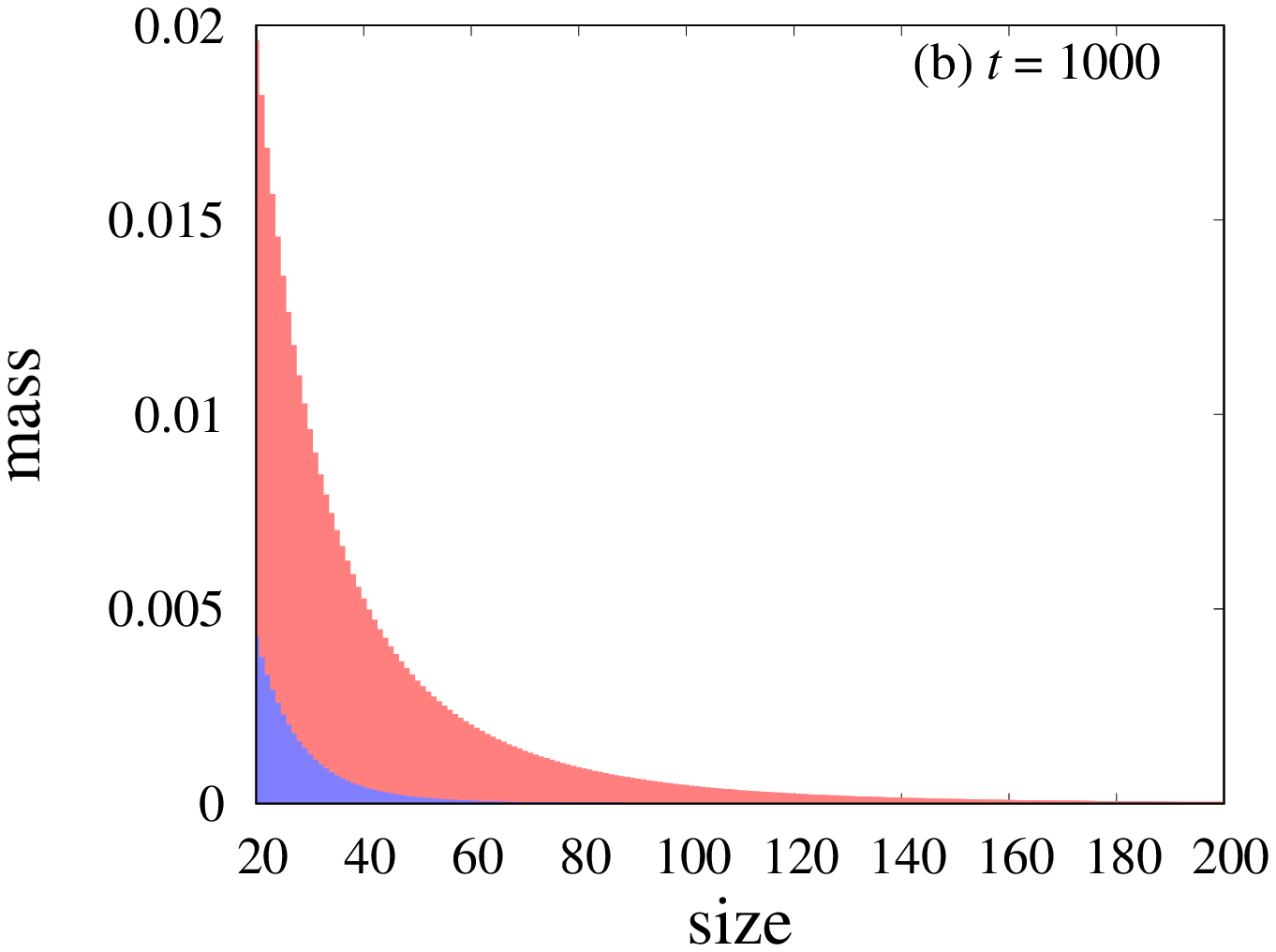}
\includegraphics[width=0.32\textwidth,clip]{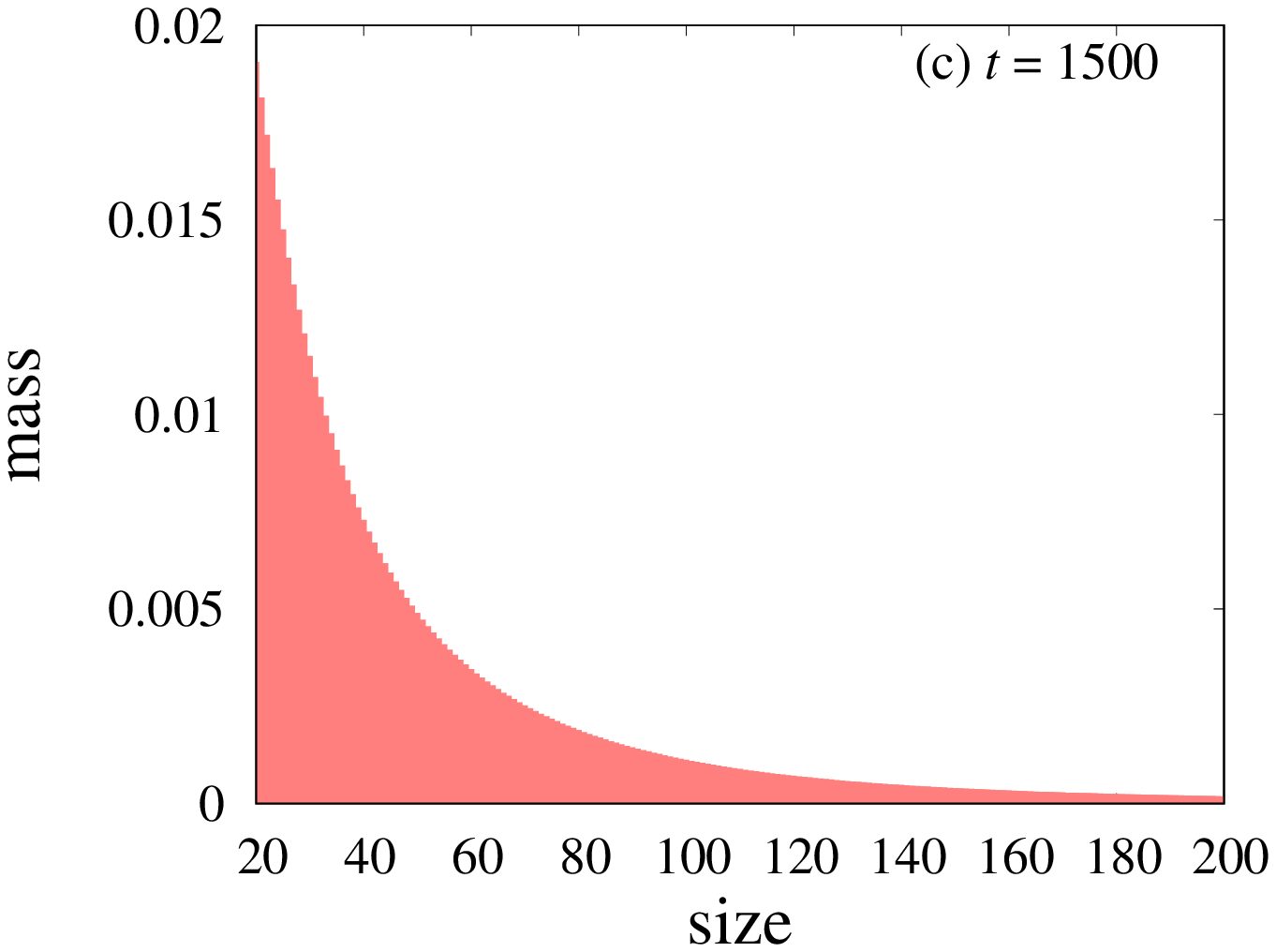}
\caption{
 The crystal mass distribution at (a) $t=500$ ($\phi=0.18$), (b) $t=1000$ ($\phi=0.65$), and (c) $t=1500$($\phi=1$).
 The initial distribution is $in_{i,\gamma} = \delta_{i,500}(1\pm\phi)/2$ with $\phi=0.05$.
 Red (light gray) and blue (dark gray) areas represent the masses of metastable crystals and stable crystals.
}
\label{fig:fortran/BDmodel/12temp/test12/param4/omega-diffneq/0.99/init/0.05/analysis/tall/dist}
\end{figure}

Fig.~\ref{fig:fortran/BDmodel/12temp/test12/param4/omega-diffneq/0.99/init/0.05/analysis/tall/dist} shows
 the crystal mass distributions $in_{i,\gamma}$ at $t=500, 1000, 1500$.
The corresponding time change of the excess parameter $\phi$ is shown with open triangles in Fig.~\ref{fig:fortran/BDmodel/12temp/test12/param4/omega-diffneq/0.99/init/0.05/ee}.
Red (light gray) and blue (dark gray) areas represent mass distributions of the metastable crystals and the stable crystals.
The initial crystal distribution is $in_{i,\rm m}=0.525\delta_{i,500}$ and $in_{i, \rm s}=0.475\delta_{i,500}$ with the initial excess parameter $\phi(0)=0.05$.
In the initial relaxation ($t \le 50$),
 the crystals dissolve rapidly due to undersaturation.
Then, the distribution takes a monotonically decreasing form at the end of high temperature period
as shown in Fig.~\ref{fig:fortran/BDmodel/12temp/test12/param4/omega-diffneq/0.99/init/0.05/analysis/tall/dist}(a).
Thereafter, the mass difference increases as the whole distribution gradually spreads [Fig.~\ref{fig:fortran/BDmodel/12temp/test12/param4/omega-diffneq/0.99/init/0.05/analysis/tall/dist}(b)].
Finally, the metastable crystals dominate the system,
 and the stable crystals disappear completely [Fig.~\ref{fig:fortran/BDmodel/12temp/test12/param4/omega-diffneq/0.99/init/0.05/analysis/tall/dist}(c)].

\begin{figure}[h]
\centering
\includegraphics[width=\linewidth]{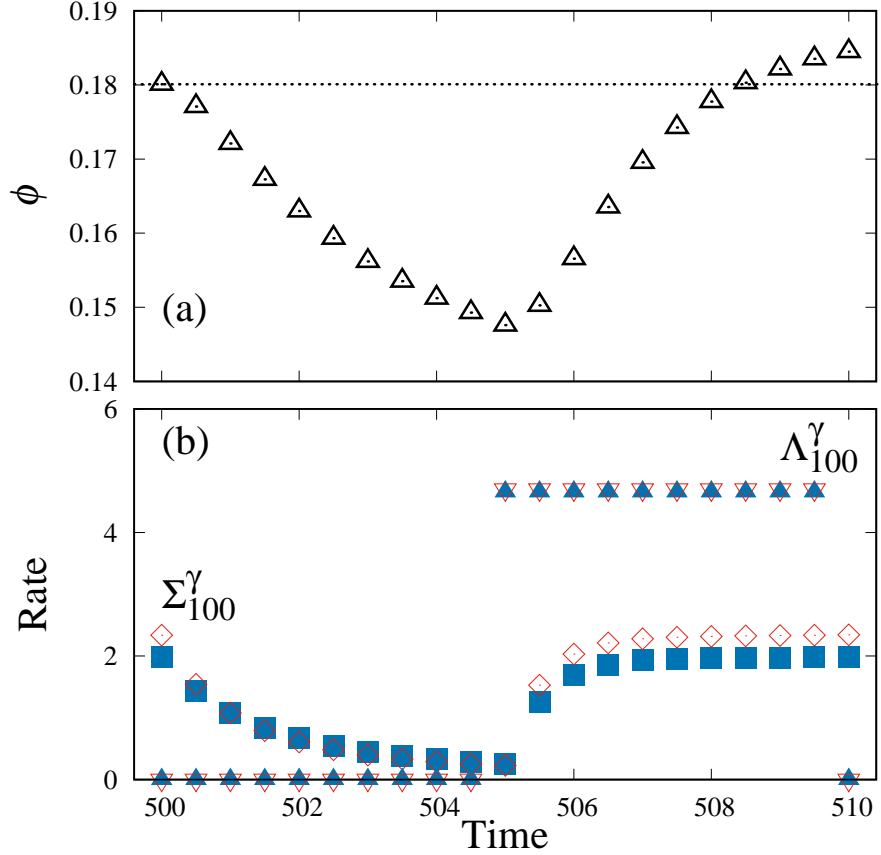}
\caption{
 Time change of (a) the excess parameter $\phi$, (b) the incorporation $\Sigma_{100}^{\gamma}$ and the dissolution $\Lambda_{100}^{\gamma}$ of a crystal of the size $i=100$ [see Eq.~(\ref{eqn:vel})] in the 51st period.
 The low temperature period is $500\leq t <505$, and the high temperature period is $505 \leq t <510$.
 The dotted line shows the magnitude of the excess parameter at $t=500$.
 Open diamonds and filled squares represent $\Sigma_{100}^{\gamma}$ for the metastable crystals and the stable crystals, respectively.
 Open inverted triangles and filled triangles represent $\Lambda_{100}^{\gamma}$ for a metastable crystal and a stable crystal, respectively.
}
\label{fig:fortran/BDmodel/12temp/test12/param4/omega-diffneq/0.99/init/0.05/analysis/tall/ees}
\end{figure}

To investigate the process in one temperature cycle,
 the time change of the excess parameter in the 51st period is plotted in Fig.~\ref{fig:fortran/BDmodel/12temp/test12/param4/omega-diffneq/0.99/init/0.05/analysis/tall/ees}(a).
The low temperature period is $500\leq t <505$, and the high temperature period is $505 \leq t <510$.
The dotted line shows the magnitude of the excess parameter $\phi$ at the beginning of the 51st period,
 and, at the end of the cycle, $\phi$ slightly increases compared to the beginning.
Both type of crystals grow thanks to the supersaturation of monomers/clusters which have been supplied in the high temperature period.
It is necessary to calculate the supersaturation in order to determine which crystal is more favorable for growth.
However, the conventional definition of supersaturation using the number of monomers is inadequate
 because of the presence of cluster incorporation.
Instead, we use the growth rate of a given size crystal as the degree of saturation.
The growth rate $v_{i}^{\gamma}$ of a size $i$-mer of the phase $\gamma$ is defined by the rate of incorporation/dissolution of monomers and clusters:
\begin{align}
v_{i}^{\gamma} &= \sum_{j=1}^{i_{\rm s}} j\left(\sigma_{i,j}n_{j,\gamma}-\lambda_{i,j}^{\gamma}\right) \equiv \Sigma_{i}^{\gamma} - \Lambda_{i}^{\gamma},
\label{eqn:vel}
\end{align}
where $\Sigma^{\gamma}_{i}$ and $\Lambda^{\gamma}_{i}$ represent the sum of the incorporation terms $j\sigma_{i,j}n_{j}^{\gamma}$ and the sum of the dissolution terms $j\lambda_{i,j}^{\gamma}$, respectively.
The conventional relation between the supersaturation and the growth rate is obtained when $i_{\rm s}=1$.

Figure \ref{fig:fortran/BDmodel/12temp/test12/param4/omega-diffneq/0.99/init/0.05/analysis/tall/ees} (b) shows $\Sigma_{100}^{\gamma}$ (squares and diamonds) and $\Lambda_{100}^{\gamma}$ (triangles and inverted triangles) in the same 51st period.
Open and filled symbols represent the data for the metastable and the stable phase crystals, respectively.
As the growth rate is the difference between the incorporation $\Sigma_{100}^{\gamma}$ and the dissolution $\Lambda_{100}^{\gamma}$,
 the stable and the metastable crystals grow at low temperature ($500\leq t<505$)
 and dissolve at high temperature ($505\leq t<510$).
In the low temperature period,
 there is no significant difference in the growth rates.
In the high temperature period,
 crystals dissolve and the excess parameter increases.
Fig.~\ref{fig:fortran/BDmodel/12temp/test12/param4/omega-diffneq/0.99/init/0.05/analysis/tall/ees}(b) shows
 that the metastable crystals are more difficult to dissolve than the stable crystals
 because the incorporation contribution of the metastable crystal $\Sigma^{\rm m}_{100}$ is more than that of the stable crystal $\Sigma^{\rm s}_{100}$
 while the dissolution contributions $\Lambda_{100}^{\gamma}$ are not so different.
This is because
 the term $\Lambda^{\gamma}_{100}$ related to dissolution is independent of the cluster size distribution
 since we have assumed that the solubility of the metastable phase is not much different from that of the stable phase.
The term $\Sigma^{\gamma}_{100}$ related to growth depends on the number of monomers and clusters.
At the beginning of the high temperature period,
 the amount of the metastable clusters produced by the dissolution is larger than that of the stable clusters
 because of the abundant metastable crystals.
As a result, 
 the excess parameter $\phi$ increases at high temperature excessively compensating the loss at low temperature.

\begin{figure}[t]
\centering
\includegraphics[width=\linewidth]{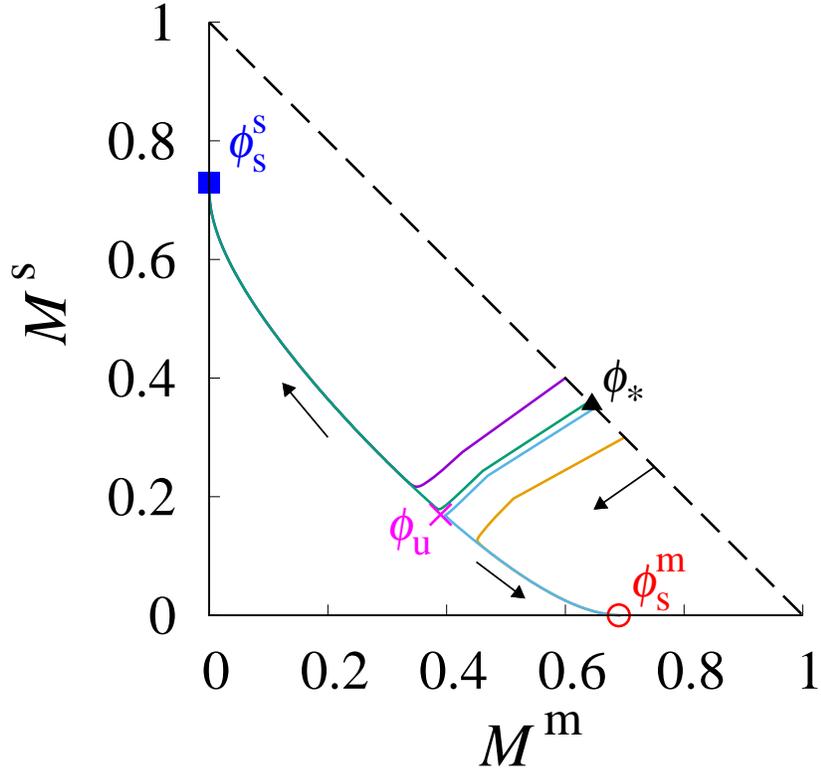}
\caption{
Flow diagram from various initial conditions ($M^{\rm m}+M^{\rm s}=1$) with $\phi(0)=0.2,0.28,0.3,0.4$
 for the solubility ratio $n_{1,\rm s}^{\rm eq}/n_{1, \rm m}^{\rm eq} = 0.9$.
The two stable fixed points are located at $(0.69,0)$ (open red circle) and $(0,0.73)$ (filled blue square).
The unstable fixed point and the critical value of the parameter are located at $(0.39,0.17)$ (magenta cross) and $(0.645, 0.355)$ (filled black triangle), respectively.
}
\label{fig:fortran/BDmodel/12temp/test12/param4/omega-diffneq/total3/flow-m-n09}
\end{figure}
\begin{figure}[t]
\centering
\includegraphics[width=\linewidth]{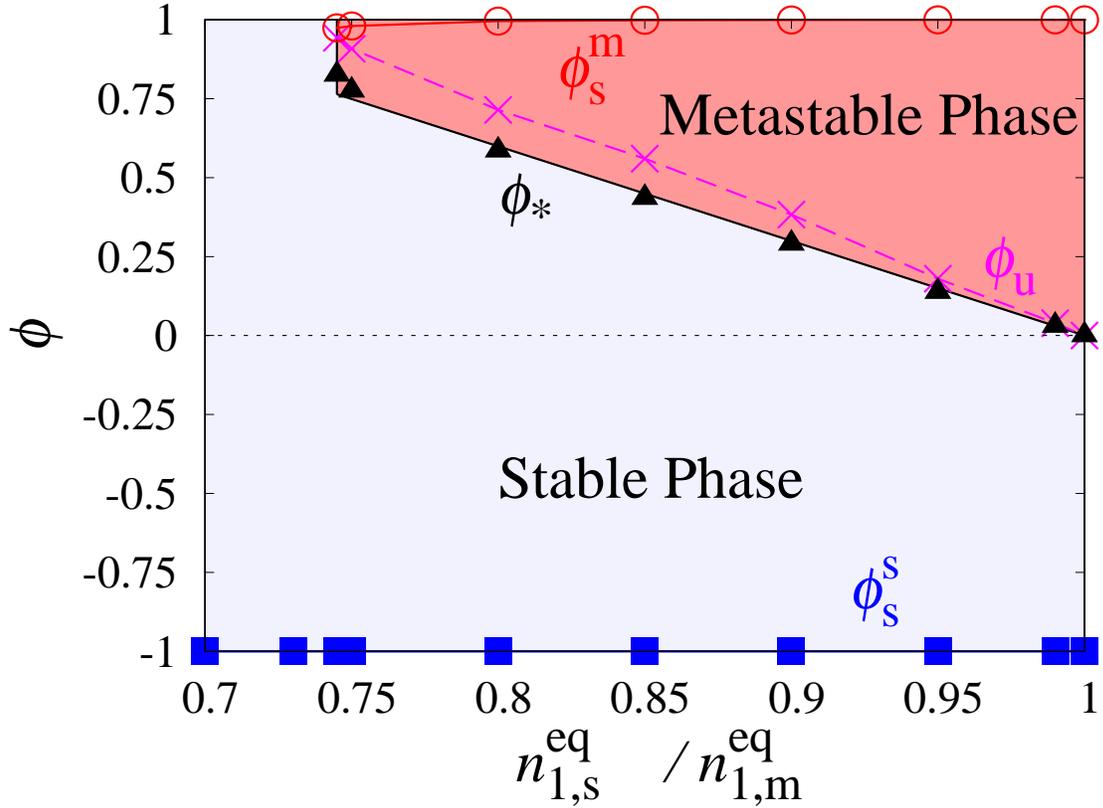}
\caption{
The phase conversion diagram shows the final phase for the given solubility ratio and the initial excess parameter $\phi(0)$.
In the red area and the blue area, the final phase is metastable and stable, respectively.
Black triangles, open circles, blue squares and yellow crosses represent
 $\phi_{\ast}$, $\phi^{\rm m}_{\rm s}$, $\phi_{\rm s}^{\rm s}$ and $\phi_{\rm u}$, respectively.
Lines are guides for the eyes.
}
\label{fig:fortran/BDmodel/12temp/test12/param4/omega-diffneq/total3/mass-g-us-bp_v2}
\end{figure}

We search the parameter range where the energetically unfavorable metastable phase can win the stable phase in TC.
Figure \ref{fig:fortran/BDmodel/12temp/test12/param4/omega-diffneq/total3/flow-m-n09} shows the flow diagram in a mass space
 at the beginning of low temperature period
 for the solubility ratio $0.9$.
In our initial condition, 
 all molecules are in crystals: $M^{\rm m} + M^{\rm s}=1$.
When the TC starts,
 both types of crystals dissolve and the masses of crystals decrease.
After a rapid initial relaxation,
 passing by either side of the unstable fixed point,
 the system reaches one of the stable fixed points.
From the two flow lines of $\phi(0)=0.28$ and $0.30$, 
 the critical value of the parameter $\phi_{\ast}=0.29$ with respect to the initial condition is found.
If $\phi(0) > \phi_{\ast}$, 
 all stable crystals disappear and only metastable crystals remain.
The final state is characterized by the excess parameter $\phi(\infty) \equiv \phi^{\rm m}_{\rm s} = 1$.
If $\phi(0) < \phi_{\ast}$, 
 the metastable crystals disappear and the stable crystals remain, that is, $\phi(\infty)\equiv \phi_{\rm s}^{\rm s} = -1$.
The crystal mass at the stable fixed point of the stable phase is larger than that of the metastable phase owing to the solubility.
The two flow lines with $\phi(0)=0.28$ and $0.30$, are parallel during the initial relaxation,
 but their directions become opposite at the unstable fixed point $(0.39, 0.17)$. 
The unstable fixed point in the mass space is characterized by the excess parameter $\phi_{\rm u}=0.39$.

From the flow diagram for various solubility ratios,
 the phase conversion diagram is obtained as shown in Fig.~\ref{fig:fortran/BDmodel/12temp/test12/param4/omega-diffneq/total3/mass-g-us-bp_v2}.
Black triangles show the critical value $\phi_{\ast}$ in the initial condition.
The critical value $\phi_{\ast}$ increases linearly on decreasing the solubility ratio.
However, the critical value $\phi_{\ast}$ vanishes around the solubility ratio $n_{\rm 1,s}^{\rm eq}/n_{\rm 1,m}^{\rm eq}\simeq 0.74$
 because the stable fixed point $\phi^{\rm m}_{\rm s}$ and the unstable fixed point $\phi_{\rm u}$ disappear.
The stable crystals prevail in the final state starting from any initial condition if the solubility ratio is less than the value: $n_{\rm 1,s}^{\rm eq}/n_{\rm 1,m}^{\rm eq} \le  0.74$,
 because the stable fixed point related to the stable phase crystal $\phi_{\rm s}^{\rm s}$ is always present.
This behavior of $\phi$ is the subcritical pitchfork bifurcation.

\section{Summary}
We investigated the conversion from the stable phase crystals to the
 metastable phase crystals by simple temperature cycling with the use of the generalized BD model.
During TC, crystals grow at low temperature and dissolve at high temperature.
When crystals dissolve, the majority metastable crystals yield a larger amount of clusters,
 which prevent the dissolution of the majority crystals through the large incorporation term $\Sigma_{i}^{\gamma}$.
At the end of one cycle, the excess parameter is amplified.
From the flow diagram in the mass space for various solubility ratios,
 the phase conversion diagram is constructed.
If the solubility ratio $n_{\rm 1,s}^{\rm eq}/n_{\rm 1,m}^{\rm eq}$ decreases,
 the initial relative amount of the metastable crystals required for the metastable phase to become dominant is large.
The critical initial excess parameter $\phi_{\ast}$ increases linearly % with the solubility ratio
 and makes a jump to $\phi_{\ast}=1$
 because the stable fixed point related to the metastable phase crystals vanishes.
The behavior of $\phi$ corresponds to the subcritical pitch fork bifurcation.

To the best of our knowledge, there is no experimental result that directly matches our scheme.
We believe that the experimental verification would be possible if the difference in solubility between stable and metastable crystals is small.
Although the VR+TC experiment of aspartic acid crystals has succeeded in converting stable racemic crystals into metastable chiral crystals\cite{viedma2021new},
 the experimental system is different from our model due to the molecular compositions of the stable and the metastable phase crystals.
Theoretical study of the phase conversion of aspartic acid crystals is now underway.

%\acknowledgments
\begin{acknowledgments}
This work is supported by JSPS KAKENHI Grant Number JP18K03500 and JP21K03379.
\end{acknowledgments}

\end{document}